\documentclass[conference]{IEEEtran}
\IEEEoverridecommandlockouts
\usepackage{cite}
\usepackage{amsmath,amssymb,amsfonts}
\usepackage{algorithmic}
\usepackage{graphicx}
\usepackage{wrapfig}  
\usepackage{textcomp}
\usepackage{subfigure}
\usepackage{multirow}
\usepackage[table,xcdraw]{xcolor}
\usepackage{bm}
\usepackage{fancyhdr}
\def\BibTeX{{\rm B\kern-.05em{\sc i\kern-.025em b}\kern-.08em
    T\kern-.1667em\lower.7ex\hbox{E}\kern-.125emX}}

\fancypagestyle{firstpage}{
  \fancyhead[C]{This paper is published in 2024 Design, Automation and Test in Europe Conference (DATE'24) Late Breaking Results} 
  \fancyfoot[C]{}
}

\begin{document}

\title{
PIMSIM-NN: An ISA-based Simulation Framework for Processing-in-Memory Accelerators
\vspace{-10pt}}
\author{
    \IEEEauthorblockN{Xinyu Wang\IEEEauthorrefmark{1}\IEEEauthorrefmark{2}, Xiaotian Sun\IEEEauthorrefmark{1}\IEEEauthorrefmark{2}, Yinhe Han\IEEEauthorrefmark{1}, Xiaoming Chen\IEEEauthorrefmark{1}\IEEEauthorrefmark{3}}
    \IEEEauthorblockA{\IEEEauthorrefmark{1}Institute of Computing Technology, Chinese Academy of Sciences; \IEEEauthorrefmark{2}University of Chinese Academy of Sciences}
    \IEEEauthorblockA{\IEEEauthorrefmark{3}Corresponding author (e-mail: chenxiaoming@ict.ac.cn)}\vspace{-30pt}
} 

\maketitle
\thispagestyle{firstpage}

\begin{abstract} 
Processing-in-memory (PIM) has shown extraordinary potential in accelerating neural networks.
To evaluate the performance of PIM accelerators, we present an ISA-based simulation framework including a dedicated ISA targeting neural networks running on PIM architectures, a compiler, and a cycle-accurate configurable simulator.
Compared with prior works, this work decouples software algorithms and hardware architectures through the proposed ISA, providing a more convenient way to evaluate the effectiveness of software/hardware optimizations.
The simulator adopts an event-driven simulation approach and has better support for hardware parallelism.
The framework is open-sourced at https://github.com/wangxy-2000/pimsim-nn.
\end{abstract}

\begin{IEEEkeywords}
Processing-in-memory, neural network accelerator, simulator, instruction set architecture
\end{IEEEkeywords}

\vspace{-10pt}
\section{Introduction} 
\vspace{-4pt}

Deep neural networks (DNNs) have shown remarkable performance in various fields, which require a large number of matrix-vector multiplications (MVMULs). Memristor-based processing-in-memory (PIM) architectures are efficient for MVMULs, exhibiting great potential in building high-throughput and low-power DNN inference accelerators.

Previous works proposed simulators to evaluate the performance of PIM accelerators. 
However, simulators that support independent designs of software and hardware have not been fully explored. 
MNSIM2.0\cite{MNSIM} is a system-level PIM simulator. 
It uses a fixed data-path and a tightly-coupled software-hardware design, limiting the optimization space. 
For example, a PE in MNSIM2.0 can either execute pooling or MVMUL, but can not execute pooling on its MVMUL outputs directly.
PUMA\cite{PUMA} is a PIM accelerator architecture, which proposes a heterogeneous ISA including core-level and tile-level.
The heterogeneous design makes it hard for hardware implementation and software synchronization.

\begin{figure}[h] 
	\centering
    \includegraphics[width=0.9\columnwidth, keepaspectratio]{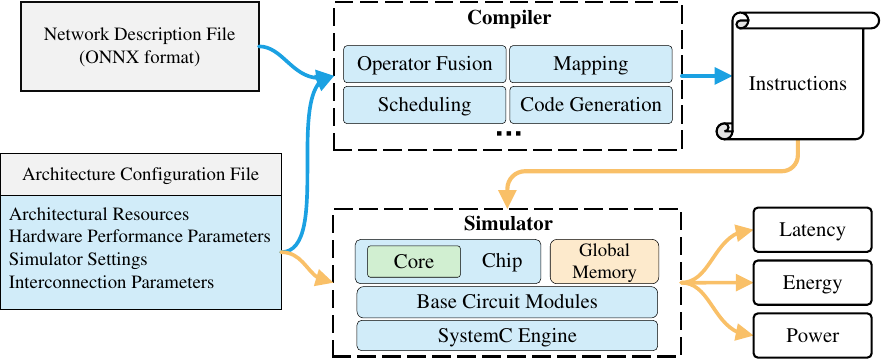}\vspace{-10pt}
	\caption{Workflow of the simulation framework.} \vspace{-10pt}
	\label{fig:workflow}
\end{figure}

In this work, we propose an instruction set architecture (ISA)-based simulation framework for crossbar-based PIM accelerators running DNNs to enable independent software optimization and hardware design space exploration. 
The framework includes an ISA, a compiler, and a simulator.
Fig.~\ref{fig:workflow} shows a brief workflow of the simulation framework.
Users first provide a architecture configuration file including hardware resources and performance parameters, and a network description file. 
Then the compiler will convert the network description into instructions considering the architecture configuration.
Finally, the simulator runs a cycle-accurate evaluation of the instructions, and shows latency, power, and energy results.

\vspace{-10pt}

\section{Instruction Set Architecture}  
\vspace{-4pt}
The ISA considers the computational requirements of DNNs and the characteristics of PIM architectures.
Most instructions in the ISA are high-level abstractions of primary operators used in DNNs, which are classified into four types: {\tt matrix}, {\tt vector}, {\tt transfer} and {\tt scalar}. 
Matrix instructions are used to control crossbars to perform efficient matrix-vector multiplications.
Transfer instructions are synchronized to simplify the hardware design.
The ISA assumes an abstract architecture: cores and global memory are connected via a sort of interconnection, and the core contains crossbars and other execution units. 
The core has a local memory, which can be accessed by {\tt matrix}, {\tt vector}, {\tt transfer} instructions.
Considering a matrix may span many crossbars, we design a group mechanism.
Crossbars belonging to the same matrix and using the same inputs will be viewed as the same group and can run in parallel.
\cite{PIM-ISA} gives more details about the ISA.


\section{Simulation Framework}
\vspace{-4pt}
\subsection{Compiler}
\vspace{-4pt}
\label{sec:compiler}
The compiler will convert a DNN description to instructions for every core with software-level optimizations.
In order to validate our work, we refer to the design of PIMCOMP\cite{PIMCOMP} and develop a similar compiler.
To showcase the impact of software optimizations, we develop two mapping algorithms based on PIMCOMP.
The first one is utilization-first, the weights of layers are mapped to cores one by one in a tight way.
For a weights matrix, if current core has enough crossbars, we map the whole matrix to the core, if not, we map part of the matrix to the core according to the available crossbars.
This may result in one core storing multiple layers' weights.
The second one is performance-first, in which we map the weights of one layer to unmapped cores, ensuring that each core only stores one layer's weights.


\begin{figure}[t]
	\centering 
    \includegraphics[width=1\columnwidth, keepaspectratio]{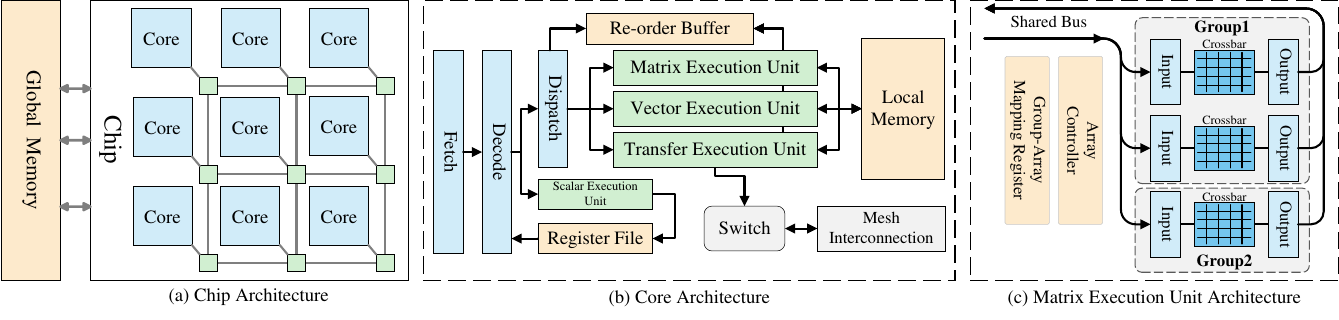}\vspace{-10pt}
	\caption{Accelerator architecture.} \vspace{-18pt}
	\label{fig:arch}
\end{figure}

\subsection{System-Level Simulator}
\vspace{-3pt}
The simulator is characterized by cycle-accurate simulation, configurability, and scalability. 
The simulator is built on the basis of {\tt SystemC}, and runs in a cycle-by-cycle way, which is good at simulating the concurrency of hardware circuits.
Users can configure their own simulator through the architecture configuration file.
The simulator is written in a modular way, allowing for convenient extension of custom modules.

The simulator assumes the accelerator using a hierarchical architecture following the proposed ISA, which is shown in Fig.~\ref{fig:arch}.
We use a mesh topology NoC as the interconnection for the chip.
The core instantiates four execution units for the four types of instructions in the ISA. 
The local memory in the core is used to store immediate results for vector/matrix operations.
The matrix execution unit contains crossbars to execute matrix-vector multiplications.
To exploit the inner parallelism, the simulator allows different execution units running in parallel with the help of the re-order order buffer (ROB) and a dispatch unit which can identify the conflicts between instructions.
\vspace{-6pt}
\section{Experimental Results}
\vspace{-1pt}
\subsection{Software/Hardware Optimization Exploration}
Our simulation framework can evaluate the optimizations of software and hardware independently.
In this part, we will show the impact of software mapping algorithms and hardware ROB capacity.
The simulator is set to a chip consisting of 64 cores, and each core has 512 crossbars, whose size is 128$\times$128, sharing with one ADC. 
\begin{figure}[b]\vspace{-18pt}
    \centering
    \subfigure[Latency]{
        \includegraphics[width=0.4\columnwidth]{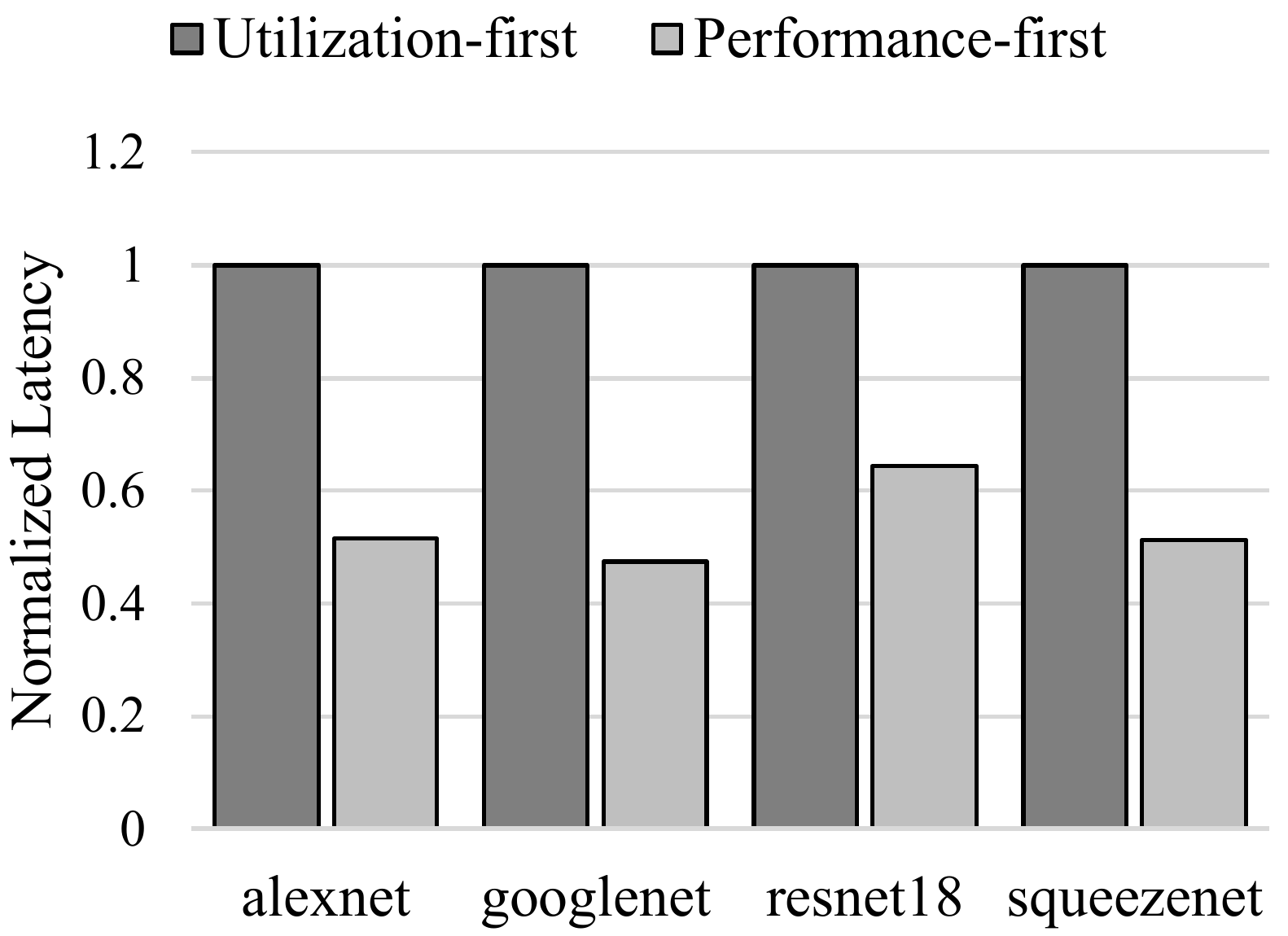}
        \label{fig:subfig1}
    }
    \subfigure[Energy]{
        \includegraphics[width=0.4\columnwidth]{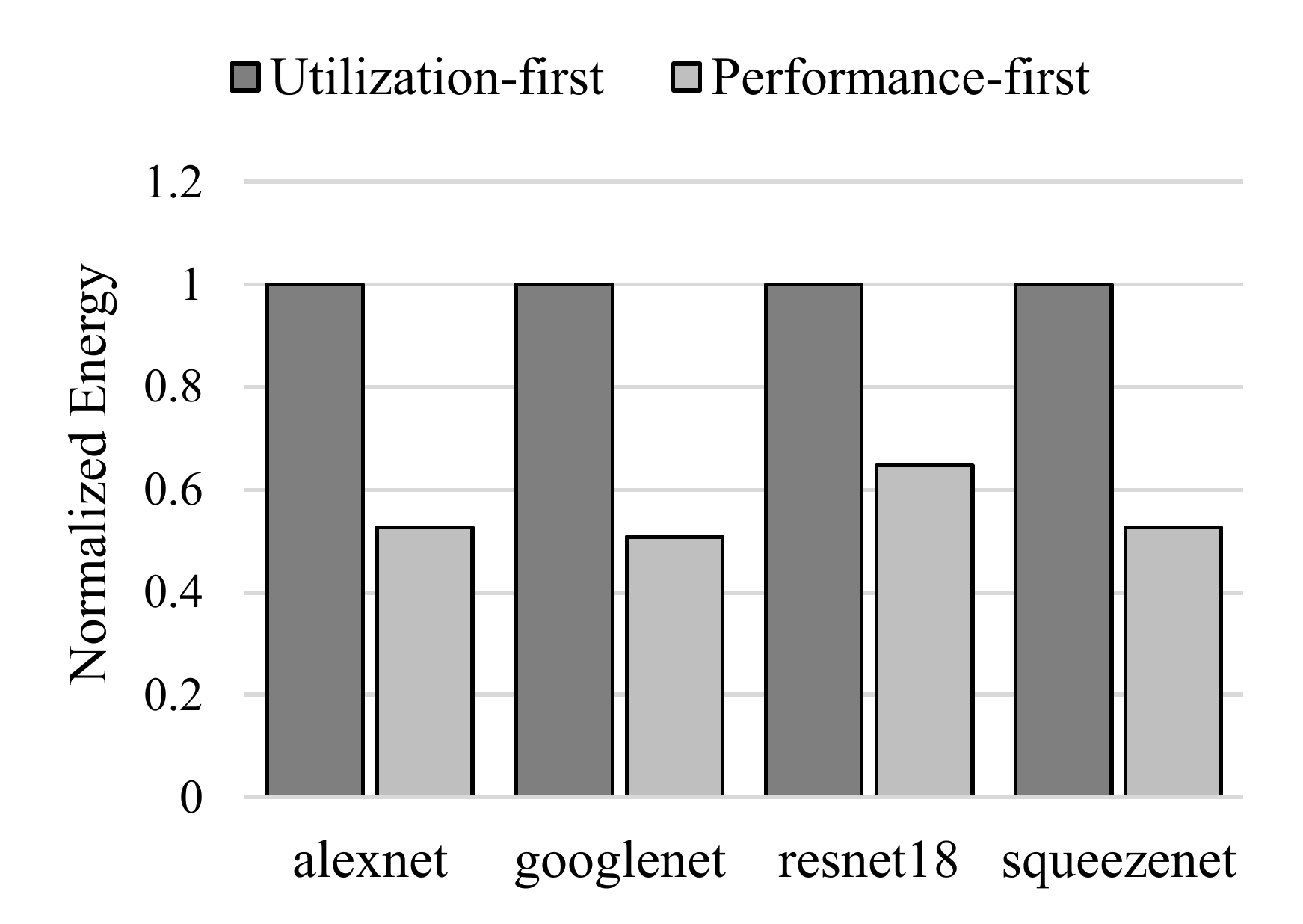}
        \label{fig:subfig2}
    }\vspace{-8pt}
    \caption{Comparison of mapping algorithms.} \vspace{-7pt}
    \label{alg_cmp}
\end{figure}

\begin{figure}[b]\vspace{-6pt}
	\centering
    \includegraphics[width=0.85\columnwidth, keepaspectratio]{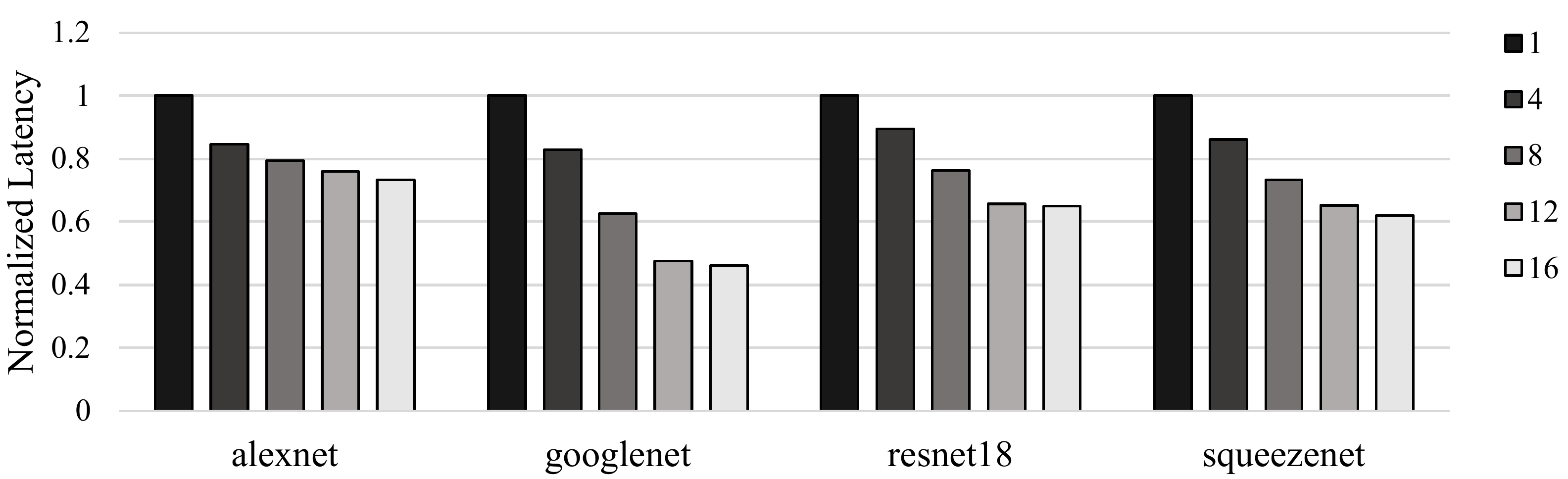}\vspace{-10pt}
	\caption{Latency results with different ROB sizes.} 
	\label{fig:rob_impact} 
\end{figure}

As mentioned in Section~\ref{sec:compiler}, we simulate the latency and energy of the utilization-first and performance-first algorithms with ROB size set to 1. 
Fig.~\ref{alg_cmp} shows the results. 
The performance-first algorithm has better latency and power than the utilization-first way, achieving 2$\times$ improvement on average, since the utilization-first one needs more intra-layer communications and reduces the parallelism for cores with multiple layers' wights.
Fig.~\ref{fig:rob_impact} shows the impact of different ROB sizes. The ROB size determines the number of instructions that can be executed simultaneously.
As the ROB size increases, the inference latency drops.
However, when we set the size from 12 to 16, the performance gain is relatively low.
This occurs when the subsequent instruction uses the same crossbar as the previous instruction in the ROB, known as structure hazard.
\vspace{-2pt}

\subsection{Comparison with MNSIM2.0}
\vspace{-2pt}
We make a comparison with a famous dataflow-based simulator MNSIM2.0 using the same crossbar configuration extracting from it.
Since its open-sourced code does not support {\tt concat} operation used in many modern DNNs, we use three modified networks in its source code to evaluate.

\begin{figure}[t]
	\centering
    \includegraphics[width=0.7\columnwidth, keepaspectratio]{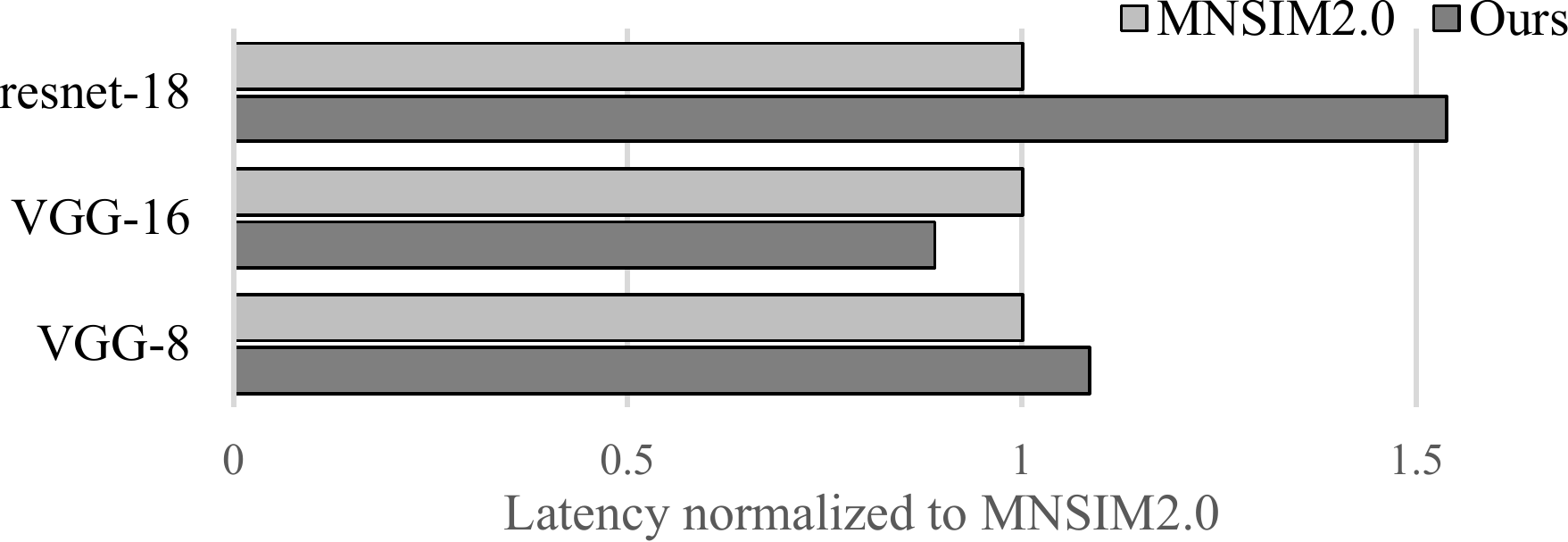}\vspace{-10pt}
	\caption{Latency comparison with MNSIM2.0.} \vspace{-18pt}
	\label{fig:compare_mnsim} 
\end{figure}

Fig.~\ref{fig:compare_mnsim} shows the latency results. For VGG networks, the latency difference between two simulators is approximately 10\%. However, our simulation on resnet-18 is 53\% slower than MNSIM2.0.
We analyse the code of MNSIM2.0 and find its data communication mechanism is overly idealistic.
It assumes fully asynchronous communication, and every data will be immediately transmitted to the next component once the data is computed.
This assumption is unrealistic as it requires an enormous buffer size and complex operation scheduling, especially for residual block, whose add operation needs to receive two layers' results.
Our ISA and simulator use synchronous communication to avoid this issue and simulate communication more accurately, while introduces some performance lost.
The communication latency ratio of the second convolutional layer, which is deeply affected by the communication, is only 18\% in MNSIM2.0, while ours is 77\%. 
\cite{Communication} indicates that the communication cost takes 40\%-90\% of the total inference latency, which matches our results. 
\vspace{-4pt}

\section{Conclusion}
\vspace{-3pt}
In this work, we present an ISA-based PIM simulation framework including an ISA, a compiler, and a simulator.
The proposal of ISA decouples software and hardware, facilitating software/hardware optimization exploration. 
Using the simulation framework, we show the impact of different optimizations on software and hardware.


\bibliographystyle{IEEEtran}
\vspace{-6pt}
\bibliography{ref}
\vspace{-1pt}
\end{document}